\documentclass[aps,twocolumn,showpacs,superscriptaddress]{revtex4-1}
\usepackage{amsmath}
\usepackage{times}
\usepackage{amssymb}
\usepackage{graphics}
\usepackage{epsfig}
\usepackage{bbm}
\usepackage{braket}
\usepackage{relsize}
\usepackage[latin1]{inputenc}

\newcommand{\States}{\mathcal{S}}
\newcommand{\Hermitian}{\mathcal{H}}

\newcommand{\tr}{\operatorname{tr}}
\newcommand{\eu}{\mathrm{e}}
\newcommand{\im}{\mathrm{i}}

\begin{document}

\title{Experimentally exploring compressed sensing quantum tomography}

\author{A.\ Steffens}
\affiliation{Dahlem Center for Complex Quantum Systems, Freie Universit\"at Berlin, 14195 Berlin, Germany}
\author{C.\,A.\ Riofr\'io}
\affiliation{Dahlem Center for Complex Quantum Systems, Freie Universit\"at Berlin, 14195 Berlin, Germany}
\author{W.\ McCutcheon}
\affiliation{Quantum Engineering Technology Laboratory, Department of Electrical and Electronic
Engineering, University of Bristol, Merchant Venturers Building, Woodland Road,
Bristol, BS8 1UB, UK}
\author{I.\ Roth}
\affiliation{Dahlem Center for Complex Quantum Systems, Freie Universit\"at Berlin, 14195 Berlin, Germany}
\author{B.\,A.\ Bell}
\affiliation{Quantum Engineering Technology Laboratory, Department of Electrical and Electronic
Engineering, University of Bristol, Merchant Venturers Building, Woodland Road,
Bristol, BS8 1UB, UK}
\author{A.\ McMillan}
\affiliation{Quantum Engineering Technology Laboratory, Department of Electrical and Electronic
Engineering, University of Bristol, Merchant Venturers Building, Woodland Road,
Bristol, BS8 1UB, UK}
\author{M.\,S.\ Tame}
\affiliation{School of Chemistry and Physics, University of KwaZulu-Natal, Durban 4001,
South Africa}
\author{J.\,G.\ Rarity}
\affiliation{Quantum Engineering Technology Laboratory, Department of Electrical and Electronic
	Engineering, University of Bristol, Merchant Venturers Building, Woodland Road,
	Bristol, BS8 1UB, UK}
\author{J.\ Eisert}
\affiliation{Dahlem Center for Complex Quantum Systems, Freie Universit\"at Berlin, 14195 Berlin, Germany}

\date{\today}

\begin{abstract}
In the light of the progress in quantum technologies, the task of verifying the correct functioning of processes and obtaining accurate 
tomographic information about quantum states becomes increasingly important. Compressed sensing,
a machinery derived from the theory of signal processing, has emerged as a feasible tool to perform robust 
and significantly more resource-economical quantum state tomography for intermediate-sized quantum systems. In this
work, we provide a comprehensive analysis of compressed sensing tomography in the regime 
in which tomographically complete data is available with reliable statistics from experimental 
observations of a multi-mode photonic architecture. Due to the fact that the data is known with high statistical significance, we 
are in a position to systematically explore the quality of reconstruction depending on the number of employed measurement settings, 
randomly selected from the complete set of data,
and on different model assumptions. We present and test a complete prescription to perform efficient compressed sensing and are able to reliably use notions of model selection and cross-validation to account for experimental imperfections and finite counting statistics. Thus, we establish compressed sensing as an effective tool for quantum state tomography, specifically suited for photonic systems.

\end{abstract}
\maketitle

\subsection*{Introduction}

Quantum technologies have seen an enormous progress in recent years. Photonic architectures have matured from basic proof-of-principle schemes
to intermediate scale quantum devices~\cite{PhotonicTechnologies}, while the robustness offered by integrated optical devices is poised to push these systems yet further~\cite{WalmsleyIntegrated,OBrienIntegrated}. 
Similarly, systems of two-digit trapped ions \cite{Blatt14} and other condensed-matter type systems  
such as superconducting devices are catching up at a remarkable pace~\cite{SuperconductingMartinis}. Building upon this technological
development, important primitives of quantum information science are being experimentally realised~\mbox{\cite{nielsenchuang,Teleportation,Bell2012,Bell2014,Bell2014a}}.
In light of these systems, it has become increasingly important to establish a toolbox for tomographic reconstruction that can keep up with this rapid development: The ironic situation that is emerging is that by now, the state of large quantum systems can be manipulated with a high degree of control, but not easily reconstructed. Clearly, these technologies and the community require further advancement of their tools for state reconstruction.  
In this work, we discuss an explicit method to achieve such a reconstruction, thus contributing to this long-term goal.
Specifically, 
we demonstrate a comprehensive exploration of the performance of state reconstruction in 
the photonic setting as one varies both the number of measurements and the noise model.

The framework of compressed sensing, a set of techniques originating from the context of classical
signal processing~\cite{CandesRombergTaoCS,CompressedSensingIntroRauhut}, has emerged as a key protagonist in closing the gap between technology and diagnostics~\cite{GrossCSPRL,Compressed2,Guta}. The idea behind its functioning is rooted in the fact that a substantial amount of data encountered in realistic situations are structured and can be characterised by significantly fewer parameters than with ad hoc schemes.
Approximately low-rank matrices are at the center of the paradigm of matrix completion in compressed sensing and correspond precisely to approximately pure quantum states.
Since pure quantum 
states are widely regarded as the key resource for quantum information processing,  such methods for reconstructing low-rank states are especially relevant.
For even larger systems,
tomographic tools based on basic variational sets are conceivable, with matrix product states~\cite{MPSTomo,Wick}, their continuous analogues~\cite{QuantumFieldTomography}, and permutationally invariant 
states~\cite{Permutation} providing prominent examples. The theory of such novel tools of reconstruction is progressing quickly. 
This applies, e.g., to new insights to the assignment of 
fair and rigorous confidence regions~\cite{Blume12,Christiandl12,Faist16,Carpentier15} as well as economical ways of performing instances of 
quantum process tomography~\cite{Compressed2,CompressedMartinis,KliKueEis15}.

Exciting steps towards using compressed sensing in experimental settings have been 
made~\cite{Permutation,CompressedMartinis,CompressedWhite,Riofrio16} in the regime in which one assumes knowledge about the basis in which sparsity is expected~\cite{CompressedMartinis}, assumes additional structure~\cite{Permutation}, or is in the highly informationally incomplete regime~\cite{Riofrio16}. In this work, we complement the picture for experimental tomography for 
medium-sized quantum systems.
In its simplest formulation, compressed sensing tomography is based on a few random expectation values of suitable observables, from which approximately low-rank states can be accurately reconstructed~\cite{GrossCSPRL}. This is suited for the situation in which expectation values can be obtained with good statistical significance, although acquiring many of them may be expensive. Still a missing piece in this 
picture, however, is the exploration of model selection techniques that have to be considered in the realm of experimental imperfections and finite counting statistics in order to make compressed sensing tomography a practical tool. 
Model selection allows to prevent over- and underfitting by controlling the dimensionality of the model of the system---in our case, the rank of the density matrix.

Here, we present a comprehensive analysis of experimental data from a multi-photon, multi-mode GHZ state source using tools of compressed sensing. Instead of working with expectation values of observables---as it is commonly done in this context, but may amount to information loss---our experimental setup allows us to obtain information on the individual projector level from the respective outcomes of each measurement setting. 
In contrast to complementing recent work \cite{Riofrio16}, we are not tied to the regime of tomographically incomplete knowledge. This allows us to study the behaviour of the reconstruction for the entire range of measurement settings.
We quantitatively explore model selection via cross validation and compare it to the model suggested by the anticipated noise statistics. With these tools, we provide a more systematic way to choose the appropriate parameters for compressed sensing quantum tomography. The results then provide the reader with the toolkit and understanding to effectively implement these methods for future quantum state tomography (QST) in general, and specifically for photonic systems. 

This work is structured as follows: We start by reviewing concepts of quantum state tomography and discuss the specifics of compressed sensing in QST. We subsequently present our experimental setup consisting of a four-qubit photonic system, which is used as a test bed for our tomographical approach. We continue by discussing concepts of model selection in the context of QST and determine the appropriate model from the experimental data. With this, we perform compressed QST and study the performance of the reconstruction depending on the amount of collected data as well as the robustness of our method with respect to model mismatches. 

\subsection*{Elements of  quantum state tomography}\label{sec:tomography}

Quantum state tomography is the most common method to diagnose quantum information processing tasks. It is used to estimate the 
unknown quantum state of a system from data produced by measuring an ensemble of identically prepared systems. By fixing a basis, a general finite-dimensional quantum state can be identified with a positive semi-definite, unit-trace matrix, the density matrix  
\begin{equation}
	\varrho\in\States_d=\{\chi\in \Hermitian_d: \chi\succeq 0, 
\tr (\chi) =1\}. 
\end{equation}
Here, $\Hermitian_d\subset \mathbb{C}^{d\times d}$ denotes the set of Hermitian matrices, and $\chi\succeq 0$ stands for a positive semi-definite matrix.

In order to determine the density matrix $\varrho$ of a quantum system, we need to prepare sufficiently many copies of the state
from identical preparations, perform a measurement on each copy using one out of $m$ different measurement settings---corresponding to different observables, i.e. Hermitian matrices $A^{(j)}$, $j=1,\dots,m$---and count the respective number of measurement outcomes. Ideal measurements are associated with unit rank projectors 
$\Pi_k^{(j)}=v_k^{(j)} v_k^{(j)\dagger}$, where $v_k^{(j)}$ is the $k$-th normalised eigenvector of $A^{(j)}$. For each measurement setting $j$ the specific 
outcome $k=1,\dots, d$ occurs with probability 
\begin{equation}
	p_{j,k}:= \tr(\Pi_k^{(j)} \varrho).
\end{equation}
Completeness, i.e. the property that the projectors sum up to unity, 
\begin{equation}
{\sum_{k=1}^d \Pi_k^{(j)}=\mathbbm{1}}, 
\end{equation}
ensures normalisation for each measurement setting $j$, so that $\sum_{k=1}^{d}p_{j,k}=1$. 
For each measurement setting $j$, the outcome $k$ corresponds to a random variable $Y_{j,k}$. 
Repeated measurements are independent from each other, and are performed on $N_j$ copies of the state for each measurement setting $j$, yielding the respective integer-valued realisation $y_{j,k}$ as observed frequency with $\sum_{k=1}^{d}y_{j,k}=N_j$.
Hence, for each measurement setting $j$, the probability of the random variables $(Y_{j,1},\dots,Y_{j,d})$ to take the configuration of measurement outcomes $(y_{j,1},\dots,y_{j,d})$ is given by 
\begin{equation}
	 \frac{N_j!}{y_{j,1}!\cdots y_{j,d}!}\ p_{j,1}^{y_{j,1}}\cdots p_{j,d}^{y_{j,d}} ,
\end{equation}	
following a multinomial distribution $\mathcal{M}(N_j,(p_{j,1},\dots,p_{j,d}))$.
Accordingly, we will obtain the $k$-th outcome $N_j \,p_{j,k}$ times  in expectation. 
We formalise the measurement process by introducing the linear operator 
\begin{equation}\label{eq:meas_op}
\mathcal{A}:\ \varrho\mapsto\big(N_j\tr(\Pi_k^{(j)} \varrho)\big)_{j,k}\,,
\end{equation}
which maps density matrices in $\States_d$ to matrices in~$\mathbb{R}^{m\times d}_+$, corresponding to measurement outcomes ${k=1,\dots,d}$ for different measurement settings ${j=1,\dots,m}$. 
We emphasise that $\mathcal{A}(\varrho)$ is not an experimental data matrix itself; 
according to the law of large numbers, the frequencies in each measurement realisation $\mathcal{Y}:=(y_{j,k})\in\mathbb{N}^{m\times d}$ from the experiment will converge to $\mathcal{A}(\varrho)$ with growing number of measurements~
 $N_j$, i.e. the expectation value $\mathbb{E}(Y_{j,k})$ of the random variable $Y_{j,k}$ is given by
 \begin{equation}
 	\mathbb{E}(Y_{j,k})=N_j\tr(\Pi_k^{(j)} \varrho)
\end{equation}	
for each $j,k$. Apart 
from additional systematic sources of error, e.g. due to experimental imperfections, the difference between $\mathcal{Y}$ and $\mathcal{A}(\varrho)$ is due to finite counting statistics, and in many settings, this is the largest contribution to the error.

The most straightforward approach to determine $\varrho$ from $\mathcal{Y}$ would be to attempt to invert the linear system of equations
\begin{equation}\label{eq:ArhoY}
\mathcal{A}(\varrho)=\mathcal{Y}.
\end{equation}
In general, however, noise on the data $\mathcal{Y}$ would render the \emph{reconstructed} density matrix $\hat{\varrho}$ unphysical ($\hat{\varrho} \not \succeq 0$).
A generic (full rank) density matrix in $\States_d$ is determined by $d^2-1$ independent real parameters. Hence, in general, one requires at least $d^2-1$  linearly independent equations in order to solve Eq.~\eqref{eq:ArhoY}. This is also called \emph{tomographic completeness}. 
When dealing with significantly less information, specialised reconstruction techniques are important with compressed sensing being a natural choice, which we will discuss in the next section.

In our system, we will be concerned with local Pauli measurements on each subsystem of a multi-partite state.
We measure an $n$-qubit system ($d=2^n$) using $m$ different measurement settings, each of which corresponds to an $n$-qubit Pauli operator
\begin{equation}
	A^{(j)}=\bigotimes_{i=1}^n\sigma^{(j)}_i, 
\end{equation}	
$j=1,\dots,m$, with $\sigma_i^{(j)}\in\{\sigma_x,\sigma_y,\sigma_z\}$, where $\sigma_x,\sigma_y,\sigma_z$ are the Pauli matrices. 
This is often referred to as Pauli basis measurement. The projectors of the two-qubit operator ${A^{(1)}:=\sigma_z\otimes\sigma_z}$, for example, are $\Pi^{(1)}_1=\ket{0, 0}\!\bra{0, 0}$, $\Pi^{(1)}_2=\ket{0 ,1}\!\bra{0 ,1}$, $\Pi^{(1)}_3=\ket{1 ,0}\!\bra{1 ,0}$, and $\Pi^{(1)}_4=\ket{1 ,1}\!\bra{1 ,1}$. 
For $n$ qubits, there exist $m_\mathrm{max}:=3^n$ different Pauli words in total 
(excluding the identity matrix for each qubit), 
each with $2^n$ eigenvectors, which corresponds to a maximum of $3^n\cdot 2^n$ equations in Eq. \eqref{eq:ArhoY}.  
Each set of Pauli projectors $\{\Pi_k^{(j')}\}_{k=1}^d$ for fixed setting $j'$ contains a subset of elements that is linearly independent from the projectors for all other settings. 
Hence, any number of smaller than $m_\text{max}$ measurement settings will lead to the loss of tomographic completeness. 
When performing QST on large systems, however, 
it is of practical necessity to employ as few measurement settings as possible (and often also only few repetitions per measurement setting). The key question arising in this context, therefore, is whether it is feasible to reconstruct an unknown state $\varrho$ with not only $m<m_\text{max}$ measurement settings, but a 
significantly smaller subset. The need for minimising the number of measurement settings is particularly pressing in architectures
such as linear optical ones, since high repetition rates and good statistics are available, while it can be 
tedious or costly to alter the measurement setting.
This is indeed the case in many practically relevant situations using compressed sensing schemes, which will be discussed in the next section.

\subsection*{Compressed sensing for quantum state tomography}\label{sec:CS}

By parameter counting, a state with rank $r<d$ can be completely characterized by fewer than $d^2$ parameters, that is $\sim rd$. 
However, it is  far from obvious how to acquire these parameters using fewer measurement settings and how to do so in a robust 
fashion---this is the starting point for compressed sensing~\cite{CandesRombergTaoCS,DonohoCS}. Originally conceived for reconstructing sparse vectors, the concept was extended to the recovery of low-rank matrices~\cite{CandesRecht09,CandesTao10} and adapted to the problem of QST~\cite{GrossCSPRL,GrossCSIEEE}.
Here, one again considers structured problems in which one can exploit the fact that in many useful settings approximately low rank states are of interest. This is a reasonable assumption, since most quantum information experiments aim at preparing pure states.

In order to obtain a general complex-valued low-rank matrix from measurements $\mathcal{A}$, na{\"i}vely, one would search within the set of low-rank matrices for the one that matches the measurement constraint, solving
\begin{equation}
\min_{\chi\in\mathbb{C}^{d\times d}}~ \text{rank}(\chi) ~~\text{ s.t. }~~  \mathcal{A}(\chi)=\mathcal{Y}.
\end{equation}
The key idea for compressed sensing in matrix recovery is to relax this NP-hard problem~\cite{VandenbergheBoydSemidef} into the closest convex optimisation problem~\cite{Fazel02}
\begin{equation}\label{eq:nucnorm_min_strict}
\min_{\chi\in\mathbb{C}^{d\times d}} \, \Vert \chi \Vert_{\ast} ~~\text{ s.t. }~~  \mathcal{A}(\chi)=\mathcal{Y}.
\end{equation}
We denote the nuclear norm (better known as the trace norm in the context of 
reconstructions in quantum mechanics) of a matrix $\chi$ by $\Vert \chi \Vert_{\ast}:=\tr(\sqrt{\chi^\dagger \chi})$. 
Such problems are well known to be efficiently solvable~\cite{boyd2004convex}.

The crucial question in compressed sensing is how many measurements are required to satisfactorily reconstruct the sought-after matrix. Many proofs rely on randomized measurements schemes: In Ref.~\cite{Tropp15}, it has been shown that for a general map $\mathcal{A}:\mathbb{R}^{d\times d}\rightarrow\mathbb{R}^M$ with Gaussian entries, $M\gtrsim 3r(2d-r)$ copies of $\varrho$ are 
provably sufficient for the recovery of $\varrho$. Building on this and closer to our situation is the recovery guarantee presented in Ref.~\cite{KuengLowRank}, in which $M\geq c\,rd$ copies are needed with some constant $c>0$,
for $\mathcal{A}:\States_d\rightarrow\mathbb{R}^M$, 
$\varrho\mapsto(\tr(\Pi^{(j)} \varrho))_{j=1,\dots,M}$, mapping density matrices from $\States_d$ to vectors in $\mathbb{R}^M$, 
with $\Pi^{(j)}=v^{(j)} v^{(j)\dagger}$, and $v^{(j)}$ a Gaussian vector for each $j$. In practice, numerical computations outperform these theoretical bounds. However, there is a fundamental lower bound for the number of copies, $M= 4r(d-r)-1$, using a theoretically optimal POVM with $M$ elements~\cite{Heinosaari2013}. 
Note that---in the mindset of measurement settings and outcomes---the number of outcomes $k$ per measurement setting $j$ scales with the dimension of the Hilbert space $d$. Since $M$ corresponds to $m\,d$, the number of measurement settings scales just with the rank, i.e.~$m=c\,r$.

It is in general harder to prove comparable results for deterministic measurements---in our setting with $v^{(j)}$ being eigenvectors of Pauli operators. To bridge this gap, notions of partial derandomisation have been introduced, where $v^{(j)}$ are not Gaussian, but drawn from spherical designs---certain finite subsets of the $d$-dimensional complex sphere---leading to similar statements~\cite{KuengLowRank}. Spherical designs, in turn, can be related to eigenvectors of $n$-qubit Pauli operators~\cite{Kueng2015}. Apart from results on the level of expectation values~\cite{LiuPauliRip}, less has been proven for products of single-qubit eigenvectors, the setting at hand---strikingly in contrast to the great success of the procedure in practice. These results remain stable when taking noise into account.

The measured data can be written as
\begin{align}
\nonumber \mathcal{Y}&=\mathcal{A}(\varrho)+\mathcal{N}(\varrho)\\
	&=(N_j\tr(\Pi_k^{(j)} \varrho))_{j,k}+(\eta_{j,k})_{j,k},
\end{align}
with $\mathcal{N}$ and $\eta_{j,k}$ representing the noise due to finite counting statistics. For positive semi-definite matrices such as quantum states, the nuclear norm of a matrix reduces to the trace of the matrix.
Consequently, relaxing the equality constraint in Eq.~\eqref{eq:nucnorm_min_strict} and including the positivity constraint, we arrive at the semi-definite program (SDP)~\cite{VandenbergheBoydSemidef}
\begin{equation}\label{eq:opt}	
	\min_{\chi\succeq 0} \, \tr \chi ~~\text{ s.t. }~~\Vert \mathcal{A}(\chi)-\mathcal{Y}\Vert_2^2<\varepsilon,	
\end{equation}
for some yet-to-be-determined $\varepsilon>0$ and $\Vert\cdot\Vert_2$ representing the entrywise two-norm. 
This is exactly the problem we aim to solve in order to achieve efficient QST. SDPs, being convex programs, feature a rich theory, and numerical implementation is easily achievable \cite{rockafellar1997,boyd2004}.
Note that the procedure minimizes the trace, which at first sight might seem contradictory to the requirement for density matrices to have unit trace. However, the unit trace requirement is implicitly included in the data constraint since the probabilities in the map $\mathcal{A}$ are normalised. Perfect data would lead to an optimizer with trace exactly equal to one. In turn, a relaxation of this constraint leads to a relaxation of the unit trace requirement.
As a result, generically for not too small $\varepsilon$, the optimal $\chi$, denoted by $\hat{\chi}$, will be subnormalised,
due to its location on the part of the boundary of the $\varepsilon$-ball with the lowest trace.
In order to obtain a physically meaningful reconstruction $\hat{\varrho}\in\States_d$, we find in our simulations that renormalising via
\begin{equation}	
	\hat{\chi}\mapsto \hat{\varrho}:=\frac{\hat{\chi}}{\tr (\hat{\chi})}
\end{equation}
produces the highest fidelity results.
To carry out the optimisation procedure, we employ the convex optimisation solver SDPT3 4.0~ \cite{sdpt3} together with CVX~\cite{cvx}. 
For higher Hilbert space dimensions, methods like singular value thresholding~\cite{SVT} come into play, which typically are faster, but less accurate.
	
\subsection*{Experimental setup}\label{sec:setup}

\begin{figure}[h]
	\begin{centering}
		\includegraphics[width=0.95\columnwidth]{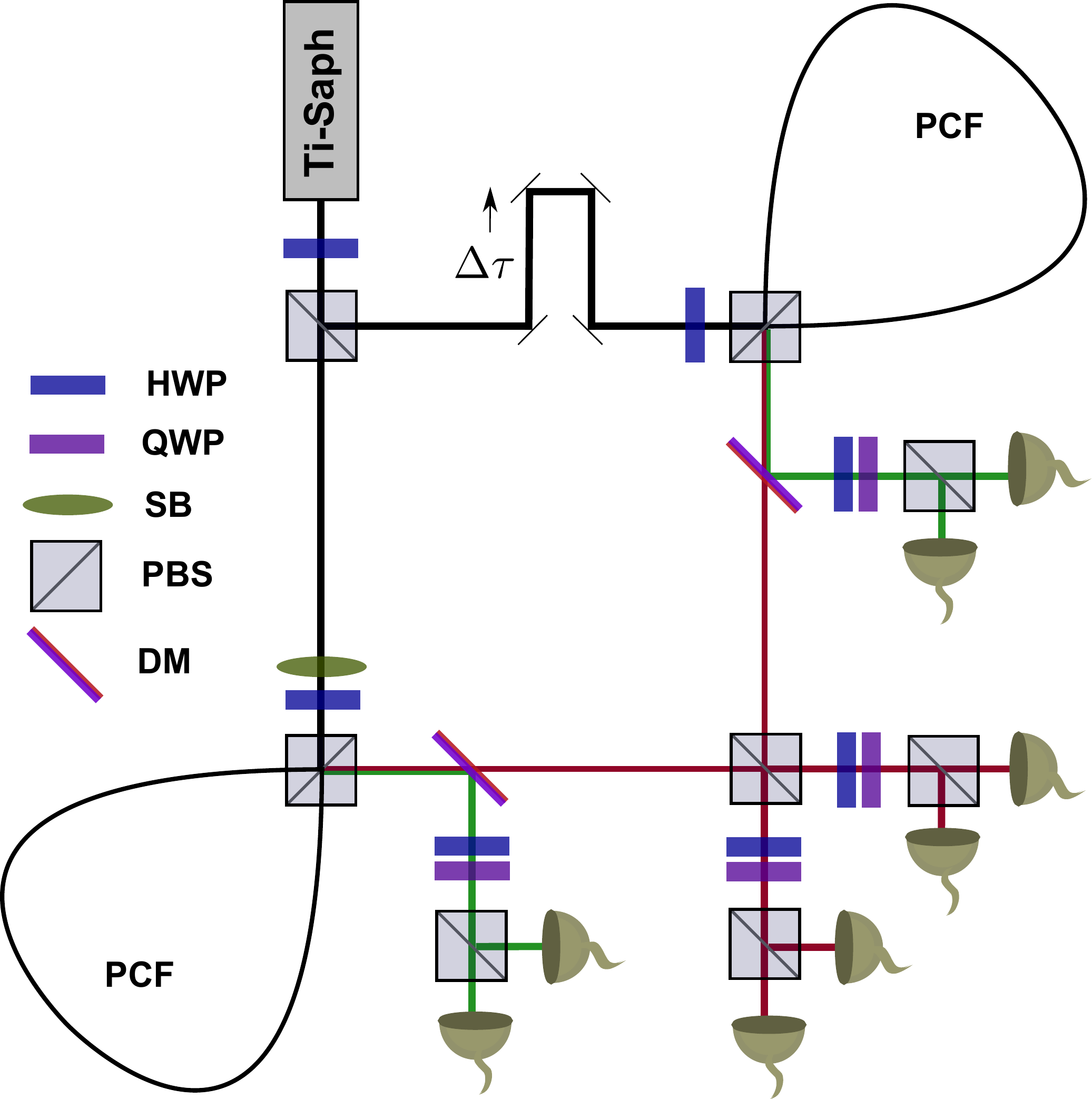}
	\end{centering}
	\caption{\label{fig:setup} Experimental setup for generating the four-photon polarisation entangled states $\ket{\psi_\mathrm{GHZ}}$, consisting of photonic crystal fiber (PCF) sources, half-wave plates (HWPs), quarter-wave plates (QWPs), a Soleil-Babinet (SB), polarising beam-splitters (PBSs), and dichroic mirrors (DMs). The 80\,MHz Ti-Saph laser is split onto two PCF sources in twisted Sagnac-loop interferometer configurations generating polarisation entangled Bell pairs. The signal and idler photons from each source are separated by DMs and the signal photons interfere on a PBS with relative time between paths $\Delta \tau\approx 0$, which on postselecting a single photon in each output port performs a fusion operation. The SB is set to match the phase between the $\ket{H,H,H,H}$ and $\ket{V,V,V,V}$ components to zero. Each mode is measured by single-qubit rotations consisting of a HWP and QWP, and is projected in the $\{\ket{H}\!,\!\ket{V}\}$-basis by PBSs and avalanche photodiode detectors. 
	}
\end{figure}

The experiment is designed to prepare the four-qubit GHZ state 
associated with the state vector
\begin{equation}
\ket{\psi_{\mathrm{GHZ}}}=\frac{1}{\sqrt2} \left(\ket{H,H,H,H}+\ket{V,V,V,V}\right)
\end{equation}
with the qubits encoded in the polarisation degree of freedom of four photons. Here, $\ket{H}$ and $\ket{V}$ represent horizontally and vertically polarized photons, respectively, 
hence effectively spanning a two-dimensional Hilbert space. The experimental 
setup, building upon the one outlined in Ref.~\cite{BellTame2013}, is shown in Fig.~\ref{fig:setup} and consists of two Bell pair sources which undergo a parity check or postselected fusion~\cite{Bell2012,Fusion,Fusion2,Pittman2001,Pan2001,Pan2003,Bodiya2006} to 
probabilistically generate the GHZ state. Both the photon pairs, generated by spontaneous four-wave-mixing in microstructured fibers, and the fusion operation are successful only probabilistically, but in a heralded fashion, i.e. a classical signal is available signifying success of the preparation. 
Successful generation of the state is determined by post-selecting only four-photon coincident events which occur at a rate of approximately $1$--$2$\,Hz. The post-selected data is effectively free from dark counts---noise generated by single photon detectors firing erroneously in the absence of a photon. This is due to the fact that the rate at which dark counts in $n$ modes occur in the coincidence window decreases exponentially with $n$, i.e. four simultaneous dark counts are negligibly rare. 
Due to additional experimental imperfections, however, the prepared state is non-ideal. The main cause of deviation between the actually 
prepared state and the target state arises from the distinguishability of photons partaking in the fusion operation and inherent mixedness from the parasitic effects in the pair generation~\cite{Bell2015}. These tend to cause the generated state to resemble a partially dephased GHZ state~\cite{Bell2012}. Measurements on the state then proceed using single qubit rotations (waveplates) and projections (polarising beam splitters and single-photon detection with avalanche photo-diodes) using well-characterised bulk-optical elements allowing high-fidelity measurements to be performed.

As stated above, in order to achieve a tomographically complete basis for $n$ qubits, one requires $m_\text{max}=3^n$ measurement settings. In our system of four qubits, $n=4$, we have measured a tomographically complete set of $81$ local Pauli operators. For each measurement setting, around $650$ four-coincident events are accumulated within an integration time of six minutes. 
Evidently, given the exponential scaling of the tomographically complete set of measurement settings, achieving such reliable statistics for larger states ($n>4$) is increasingly demanding on resources and quickly becomes infeasible.

\subsection*{Model selection}\label{sec:modelselection}

The starting point for carrying out compressed sensing quantum tomography 
is the question of determining an appropriate value for $\varepsilon$ in the optimisation procedure Eq.~\eqref{eq:opt}. 
Essentially, larger values of $\varepsilon$ result in greater relaxation of the data fitting constraint, leading to lower-rank estimates $\hat{\varrho}$; while smaller $\varepsilon$ values will yield $\hat{\varrho}$ matrices with larger rank, which better fit the particular data set. Depending on the underlying state and the particular instance of noise in the data, the choice of $\varepsilon$ might result in under-fitting with too coarse a model, or in overfitting---i.e. including parts of the noise into the model of the state. Both extremes in general lead to states that fail to correctly predict future data.
In the most severe cases, it could happen that using the same measurement prescription $\mathcal{A}$ and the same data $\mathcal{Y}$, the optimisation procedure in Eq.~\eqref{eq:opt} yields a full rank or a rank-one matrix, depending on the choice of $\varepsilon$. Worse still, too small a value of $\varepsilon$ can make the optimisation procedure unfeasible, whereby there is no feasible state that would result in data sufficiently close to that measured. The task of determining the appropriate model---in our case, the value of~$\varepsilon$---that is statistically faithful to the data via an appropriate choice of the respective external parameters is called \emph{model selection} (see e.g. Ref.~\cite{burnham2003model}). 
Several ideas of model selection have a rigorous mathematical underpinning: Particularly well-known is the Akaike information criterion (AIC)~\cite{AIC}, 
providing a measure of the relative quality of statistical models for a given set of data. 
For a collection of models compatible with a given data set, this criterion gives an estimate for the relative quality of each model. Similarly frequently employed is the 
Bayesian information criterion (BIC)~\cite{BIC}.
Direct application of AIC and BIC to quantum tomography---an approach followed in Ref.~\cite{Guta}---is
problematic for larger systems since it requires rank-restricted maximum-likelihood estimation, leading to non-convex optimisation, which scales unfavorably with the system size. This is due to the fact that these techniques are discrete in the sense that they explicitly restrict the rank of the density matrix. In the compressed sensing mindset, the parameter that controls the rank in a continuous fashion is $\varepsilon$. As we mentioned above, this is at the centre of our discussion.

For sufficiently small noise, a promising ansatz for identifying a suitable $\varepsilon$ is to use the data to compute the estimate $\hat{\varepsilon}(\mathcal{Y})$ according to the expectation value of 
\begin{equation}
\Vert \mathcal{A}(\chi)-\mathcal{Y}\Vert_2^2=\Vert\mathcal{N}(\varrho)\Vert_2^2. 
\end{equation}
Assuming the noise is solely due to finite counting statistics, i.e. the deviations from measurement outcomes from the expected variance of the multinomial distribution, we obtain
\begin{align}\label{eq:epsilon}
\nonumber\mathbb{E}(\Vert\mathcal{N}(\varrho)\Vert_2^2)=\sum_{j,k}\mathbb{E}(\eta_{j,k}^2)=\sum_{j,k}\mathbb{V}(\eta_{j,k})\\=\sum_{j,k}N_j p_{j,k}(1-p_{j,k}),
\end{align}
with variance $\mathbb{V}$. The second step follows from $\mathbb{E}(\eta_{j,k})=0$ for each $j$ and $k$. 	
In order to compute $\hat{\varepsilon}$ from the data, we need to approximate $p_{j,k}$ as ${y_{j,k}}/{N_j}$, which is reasonable for sufficiently large $N_j$ according to the law of large numbers. By Eq.~\eqref{eq:epsilon}, we obtain the estimate
\begin{align}\label{eq:hatepsilon}
\hat{\varepsilon}(\mathcal{Y}):=\sum_{j=1}^m\sum_{k=1}^{d} y_{j,k}\,(1-{y_{j,k}}/{N_j}).
\end{align}
This choice of $\hat{\varepsilon}=\hat{\varepsilon}(\mathcal{Y})$ scales linearly with $m$, the number of measurements in the dataset $\mathcal{Y}$. 
Note that $\hat{\varepsilon}$ depends on the noise model, which in several cases may not be sufficiently established. 
In our case, however, the noise model is known to a high degree, which allows us to study and compare different methods for estimating the parameter $\varepsilon$.

Complementarily, we employ a straightforward, well-established model selection technique based on cross validation (see 
e.g. Ref.\ \cite{efron1994}), which is more scalable than the use of AIC or BIC 
in our case. For this, the data is partitioned into independent training and testing sets. Different models, i.e. different values for~$\varepsilon$, are built from the training data and used to predict the testing data. The sought-after parameters---in our case $\varepsilon$---then result from the model corresponding to the smallest error with respect to the testing data.

Specifically, we randomly draw $m=10,15,20,40,60,80$ out of the $m_\text{max}=81$ measurement settings without replacement, corresponding to different levels of limited experimental knowledge. The respective data sets $\mathcal{Y}(m)\in\mathbb{R}^{m\times d}_+$ are then partitioned into five subsets $\mathcal{Z}^{(1)}(m),\dots,\mathcal{Z}^{(5)}(m)\in\mathbb{R}^{m/5\times d}_+$. The optimisation in Eq.~\eqref{eq:opt} is performed with respect to every possible union of four subsets $\bigcup_{i=1,i\neq q}^5 \mathcal{Z}^{(i)}(m)$, $q=1,\dots,5$, and different $\varepsilon$ parameters. Each reconstruction yields an estimate $\hat{\varrho}\,(m,q,\varepsilon)$ and the remaining subset  $\mathcal{Z}^{(q)}(m)$ is used as a testing set. The state estimate $\hat{\varrho}\,(m,q,\varepsilon)$ is used to compute the \emph{predicted} measurement data $\mathcal{A}_{m,q}(\hat{\varrho}\,(m,q,\varepsilon))$ and compare these with the corresponding subset of the \emph{experimental} measurement data $\mathcal{Z}^{(q)}(m)$ (${\mathcal{A}_{m,q}:\,\States_d\rightarrow \mathbb{R}^{m/5\times d}_+}$ being the reduction of the operator $\mathcal{A}$ to the subsets of measurement settings corresponding to $m$ and $q$). The resulting distance ${\Vert \mathcal{A}_{m,q}(\hat{\varrho}\,(m,q,\varepsilon))-\mathcal{Z}^{(q)}(m)\Vert_2}$, between the \emph{predicted} and \emph{measured} data, also known as the prediction error or predicted risk, is averaged over $q$ (fivefold cross-validation), yielding an estimate for the averaged prediction error (testing set error) 
\begin{equation}
	E(m,\varepsilon)=\frac{1}{5}\,\sum_{q=1}^{5}\Vert \mathcal{A}_{m,q}(\hat{\varrho}\,(m,q,\varepsilon))-\mathcal{Z}^{(q)}(m)\Vert_2.
\end{equation} 
If the corresponding optimisation problem is infeasible for a certain combination of  $\varepsilon$, $m$, and $q$ (i.e. the set of density matrices that satisfy the constraint in Eq.~\eqref{eq:opt} is empty), the prediction error is set to  $\Vert \mathcal{Z}^{(q)}(m)\Vert_2$. For averaging, each point $(m,\varepsilon)$ is sampled 50 times.

\begin{figure}[h]
\begin{centering}
\includegraphics[width=\columnwidth]{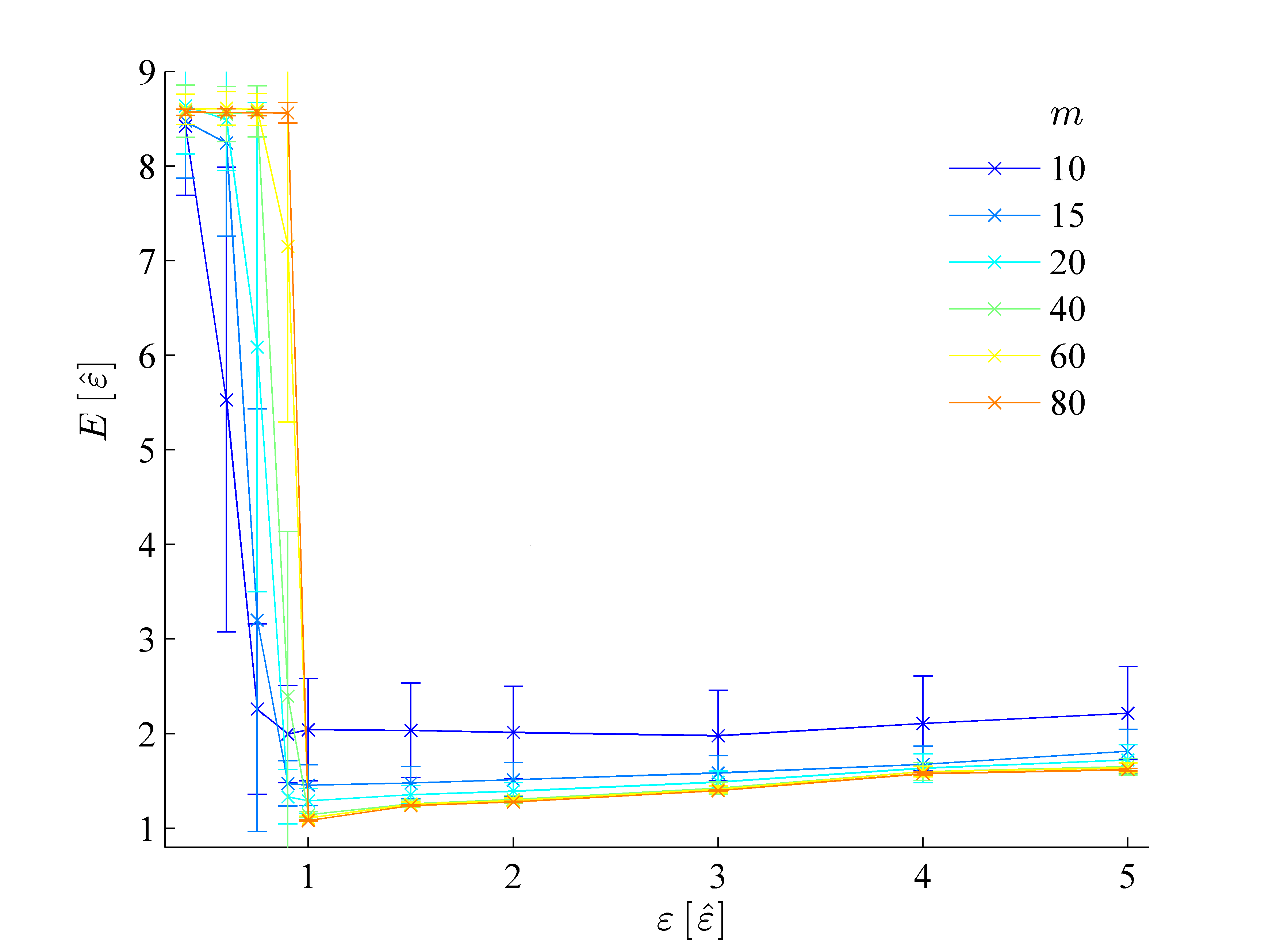}
\par\end{centering}
\caption{\label{fig:crossval}Cross validation results. Prediction errors $E(m,\varepsilon)=1/5\,\sum_{q=1}^{5}\Vert \mathcal{A}_{m,q}(\hat{\varrho}\,(m,q,\varepsilon))-\mathcal{Z}^{(q)}(m)\Vert_2$
in units of~$\hat{\varepsilon}$ depending on the model parameter $\varepsilon$ and on the number of measurement settings $m$. The standard deviation is bigger for fewer measurement settings and for smaller $\varepsilon$. The latter is due to the increasing chance of the optimisation to be infeasible for smaller $\varepsilon$. For $\varepsilon$ close to $\hat \varepsilon$ and sufficient many measurement settings, the error is only slightly bigger than the deviation due to the multinomial distribution of the measurement outcomes.}
\end{figure}

The mean values and standard deviations of the prediction error depending on the model parameter are depicted in Fig.~\ref{fig:crossval}.
We see that for values of $\varepsilon$ around $\hat \varepsilon$ the error is smallest, which is consistent with our ansatz and allows us to gain confidence in the assumption that the measurement data can be effectively modelled by a multinomial distribution. 
The more measurement settings are considered, the clearer the choice of the optimal $\varepsilon$ becomes, with both the prediction error and its variance attaining their minima close to $\varepsilon=\hat{\varepsilon}$.
For those values of $\varepsilon$ close to $\hat{\varepsilon}$ and sufficiently many measurement settings, the prediction error $E(m,\varepsilon)$ is only slightly bigger than the error estimate for the data $\varepsilon$. Here, the error arises primarily from raw multinomial noise, $\varepsilon$, present in the testing set itself and cannot be overcome with improved reconstruction methods. 
 Where fewer measurement settings are considered, less information about the state is available, resulting in large testing set errors as well as greater variance of the state estimates, although the smallest prediction errors are still seen for $\varepsilon$ close to $\hat \varepsilon$. As $\varepsilon$ decreases below $\hat \varepsilon$, the chance of the optimisation being infeasible increases, causing the prediction errors to effectively increase with a greater spread attributed to different optimisation runs. As  $\varepsilon$ increases above $\hat \varepsilon$, the data fitting constraint is weakened, resulting in too coarse model fits and a gradually increasing prediction error. 

Using Eq.~\eqref{eq:hatepsilon} instead of cross validation has the advantage of much less computational effort and is useful in a scenario with good statistics for each measurement setting. Moreover, cross validation relies on partially discarding data, which could aggravate the issues of having too little data, yielding poorer estimates for $\varepsilon$. However, Eq.~\eqref{eq:epsilon} relies on the assumption of a well identified error model---in our case, multinomial noise, as verified by cross validation. In cases in which the error model is not known, cross validation can provide a more robust estimate of $\varepsilon$. 

\subsection*{Compressed sensing tomography of the GHZ state }

\begin{figure}[h]
\begin{centering}
	\begin{minipage}[t]{\columnwidth}
		\begin{center}
			Target state{
				\includegraphics[width=0.99\columnwidth]{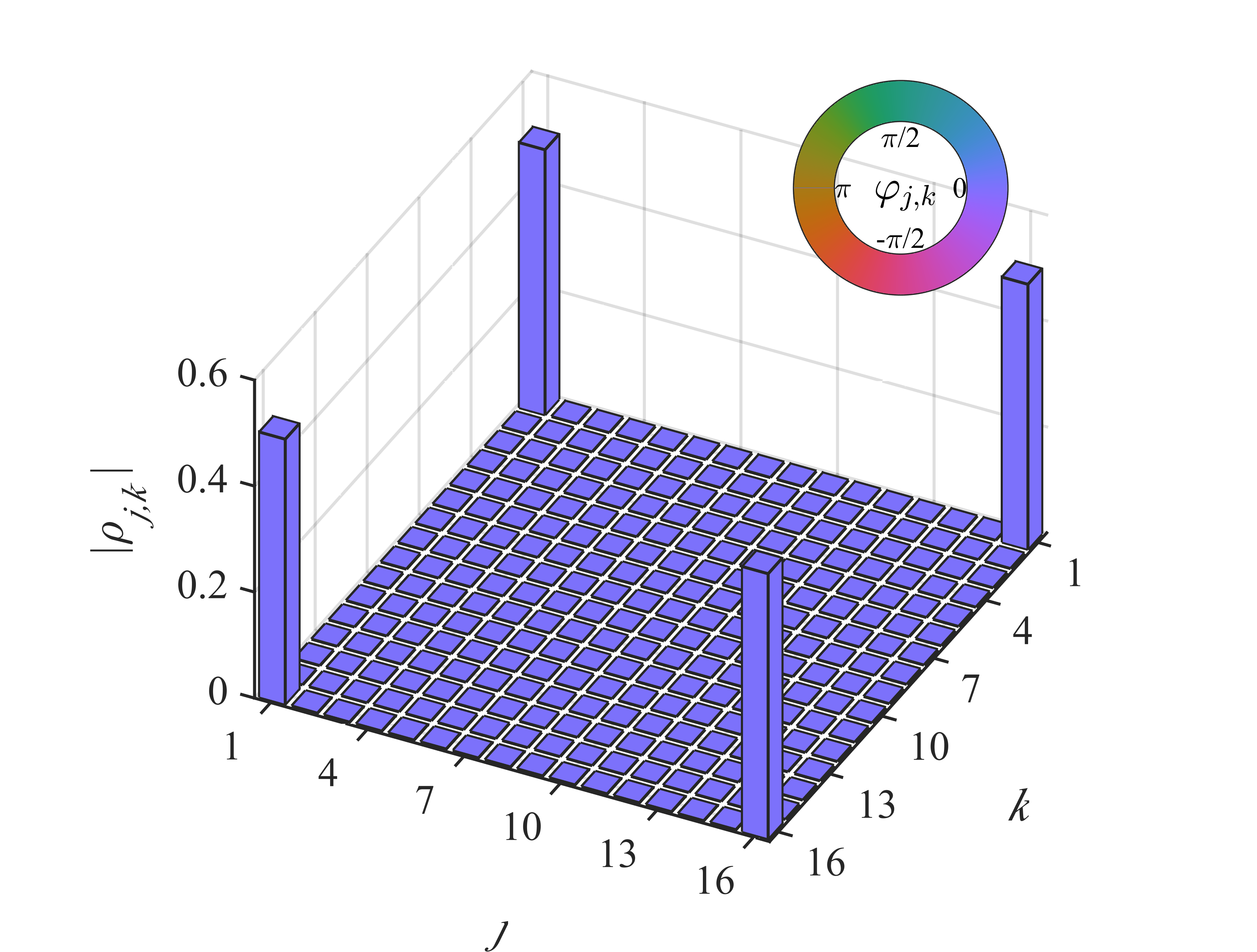}}
		\end{center}%
	\end{minipage}%
\par \vspace{.5cm}
	\begin{minipage}[t]{\columnwidth}
		\begin{center}
			Estimated state
			{
				\includegraphics[width=0.99\columnwidth]{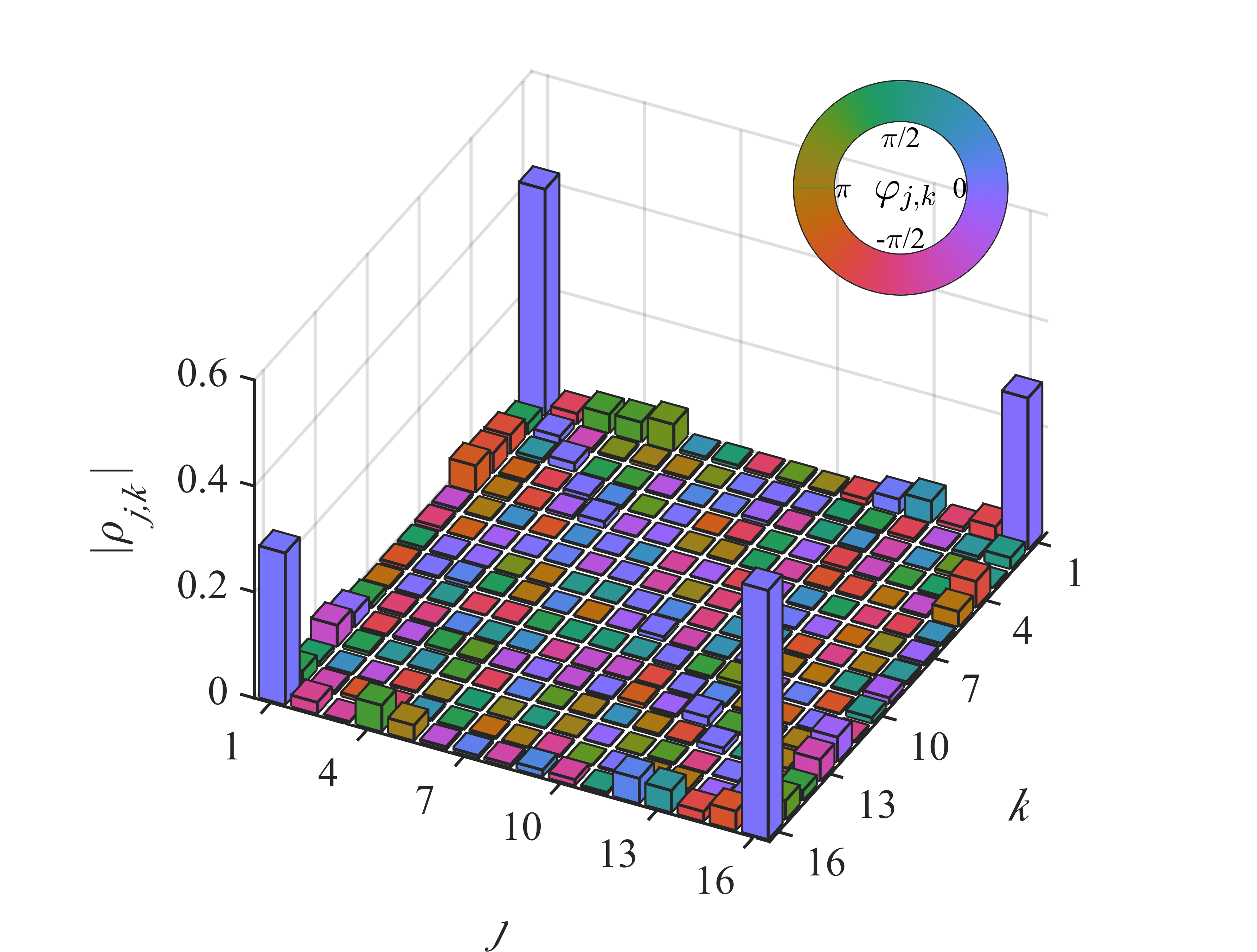}
			}
		\end{center}%
	\end{minipage}
\par
\end{centering}	
\caption{\label{fig:state} Bar plot of the density matrix of the target (GHZ) state $\rho_\mathrm{GHZ}$ and its optimal compressed sensing estimate $\hat{\rho}_\mathrm{CS}$. The basis is fixed to the tensor products of one-particle vectors in the order
$\ket{H,H,H,H},\ket{H,H,H,V},\dots,\ket{V,V,V,V}$. 
The height of each bar corresponds to the size of the absolute value of the respective density matrix entry $\varrho_{j,k}=\vert \varrho_{j,k}\vert\, \eu^{\im\varphi_{j,k}}$ and the colour to its complex phase $\varphi_{j,k}\in (-\pi,\pi]$. The colourmap is chosen to account for the periodicity of the phase. 
The fidelity of the estimate with respect to the GHZ state is $0.855\pm0.006$ and its purity $\tr(\hat \varrho_\mathrm{CS}^2)=0.60\pm0.01$, representing an expected mixedness due to experimental imperfections.}
\end{figure}

Having verified that the optimal value for $\varepsilon$ is close to that computed from Eq.~\eqref{eq:hatepsilon}, we use it as input for the compressed sensing tomography of the experimental state and compute the optimal estimate $\hat{\varrho}_\mathrm{CS}:=\hat{\varrho}\,(m_\text{max},\hat{\varepsilon})$ of the a priori unknown experimentally prepared state $\varrho$. 
The good statistics available in our 
experiment allow us to estimate $\varrho$ with comparably high accuracy. In general, due to experimental imperfections, $\varrho$ (and hence $\hat \varrho$) will deviate from the target state 
$\varrho_{\text{GHZ}} := \ket{\psi_\text{GHZ}}\!\bra{\psi_\text{GHZ}}$, see Fig.~\ref{fig:state} for a pictorial representation. There, we show a comparison between the density matrices of the target state and the optimal compressed sensing estimate using bar plots.

The standard figure of merit to determine the performance of tomography is the quantum fidelity $F$ of two states $\chi$ and $\sigma$, which is defined as  $F(\chi,\sigma)=\tr ((\sqrt{\chi}\sigma\sqrt{\chi})^{1/2})$ \cite{nielsenchuang}.
We find that the fidelity between the GHZ state $\varrho_\mathrm{GHZ}$ and the estimated state $\hat{\varrho}_\mathrm{CS}$ is 
\begin{equation}\label{eq:fidel_C}
F(\varrho_{\text{GHZ}},\hat{\varrho}_\mathrm{CS})=0.855\pm 0.006. 
\end{equation}
The uncertainty of the fidelity is determined by using the optimal compressed sensing estimate, $\hat \varrho$, as input for the generation of simulated data---parametric bootstrapping~\cite{efron1994}---and taking the empirical standard deviation of the fidelity values. This uncertainty determines the robustness of the method.
Obtaining a closed expression for proper error bounds from the data with respect to positivity constraints is hard \cite{Carpentier15,Suess16}, while bootstrapping and taking the empirical standard deviation gives a good estimate of uncertainty~\cite{efron1994}.

To build confidence, we also computed the maximum likelihood estimate~\cite{Hradil1997}, $\hat{\varrho}_{\rm MLE}$, using the same data to obtain a fidelity with respect to the target state of ${F(\varrho_{\text{GHZ}},\hat{\varrho}_{\rm MLE})=0.843\pm 0.004}$, 
which shows that the estimators yield similar results; as will other estimators such as least squares with positivity constraint.
Additionally, since we have measured a tomographically complete set of observables and the statistical properties of the measured data are sufficiently understood, we are able to provide an estimate of the fidelity with respect to the target state directly from the measured data without the need of performing tomography and an estimate of the corresponding error bound, see Appendix \ref{appx:FidelityEstimation} for details. With this, we obtain a fidelity of $0.845 \pm 0.005$,
which again is in good agreement with the results computed from the compressed sensing estimate. 
We note that the standard technique for estimating the fidelity of a state with respect to a specific target state requires estimating only the expectation values of a set of operators that form a decomposition of the target state. For a four-qubit GHZ state, this requires a minimum of nine specific Pauli basis measurements, as explained in Appendix~\ref{appx:FidelityEstimation}. In contrast, using compressed sensing tomography, even a random set of measurement settings produces fidelity estimates with respect to the GHZ state, which quickly approach the maximum at around 25~measurement settings. Furthermore, these measurement settings suffice to compute the fidelities with respect to arbitrary states, since they allow for the estimation of the entire state.

\begin{figure}[h]
	\begin{centering}
		\includegraphics[width=0.99\columnwidth]{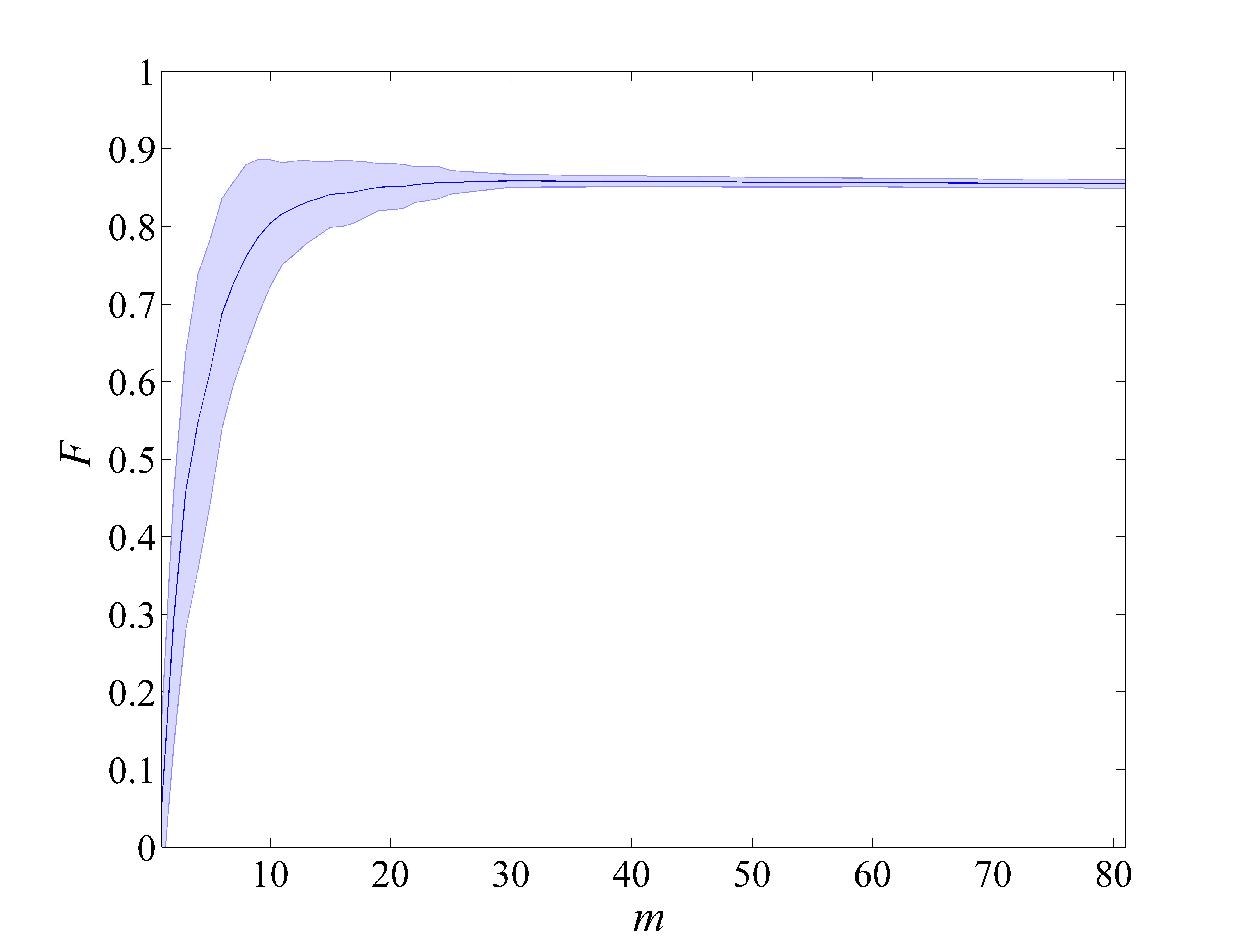}
	\end{centering}
	\caption{\label{fig:star_tomo} Fidelity $F(\varrho_{\text{GHZ}},\hat{\varrho}(m,\hat \varepsilon))$ as a function of the number of measurement settings $m$ with uncertainty (shading) from bootstrapping for $\varepsilon= \hat{\varepsilon}$. For large $m$, $F$ approaches the fidelity of $\varrho_{\text{GHZ}}$ and $\hat{\varrho}$, $F(\varrho_{\text{GHZ}},\hat{\varrho}_\mathrm{CS})=0.855$, getting very close already for comparibly few measurement settings, and the standard deviation becomes smaller.}
\end{figure}

Compressed sensing is about employing provably fewer measurement settings than with standard methods, while still producing satisfactory reconstructions, i.e. to effectively sense in a compressive way.
Along these lines, we explore how varying the number of measurement settings $m$ affects the fidelity. This is shown in Fig.~\ref{fig:star_tomo}. In order to make the results independent from specific measurement settings, we randomly draw without replacement $m$ out of $m_\mathrm{max}$ different settings 200 times and average over the resulting fidelities, thus providing a value for a typically expected fidelity for each $m$. 
As one would expect intuitively, we can see that the value of the fidelity increases monotonically with the number of measurement settings and converges to the fidelity of the estimate from tomographically complete data. The shaded region represents the uncertainty ($\pm$ standard deviation) in the fidelity computed via bootstrapping and displays the decreasing uncertainty with increasing numbers of measurement settings. The fidelity already falls within the errorbars of its final value for comparably small $m$.

\subsection*{Deviations from the optimal parameter}

In this section, we study the effect that misestimating $\varepsilon$ has in the performance of the reconstruction of the state. We carry out this task by numerical simulation: Using the compressed sensing state estimate $\hat{\varrho}_\mathrm{CS}$, we simulate measurement data, which we subsequently input to our compressed sensing reconstruction procedure, varying both $\varepsilon$, $m$ and randomly drawing measurement settings without replacement. If the corresponding optimisation problem is infeasible and yields no estimate, 
the fidelity $F$ is set to zero. The fidelities $F(\hat{\varrho}_\mathrm{CS},\hat{\varrho}(m,\varepsilon))$ are averaged over data and measurement settings (500 different data sets and different measurement settings per $m$ and $\varepsilon$).

\begin{figure}[h]
\begin{minipage}[t]{0.99\columnwidth}%
\begin{center} \label{fig:cs_F}{
\includegraphics[width=0.99\columnwidth]{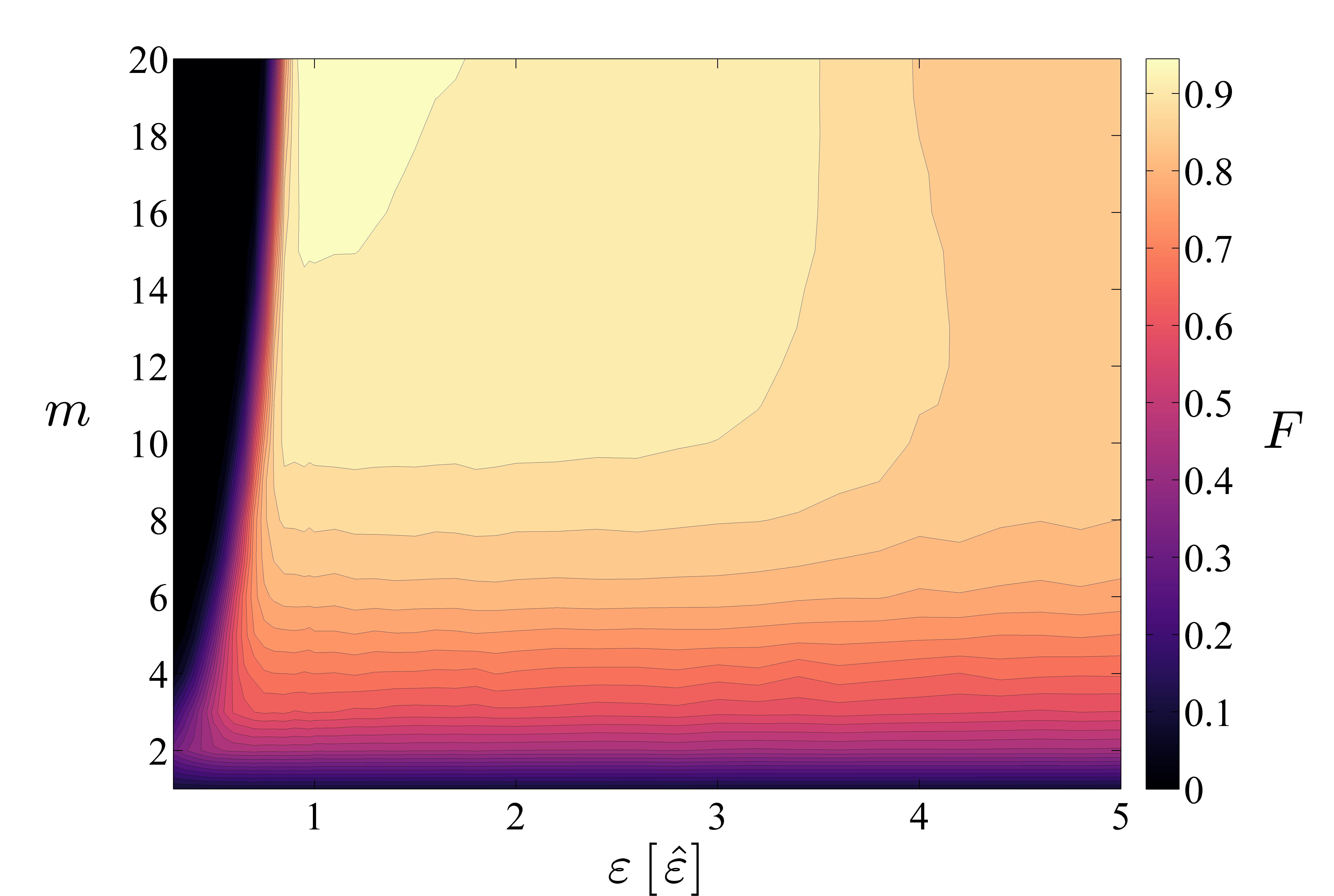}}
\end{center}
\end{minipage}
\begin{minipage}[t]{0.99\columnwidth}%
\begin{center}\label{fig:cs_std}
{
\includegraphics[width=0.99\columnwidth]{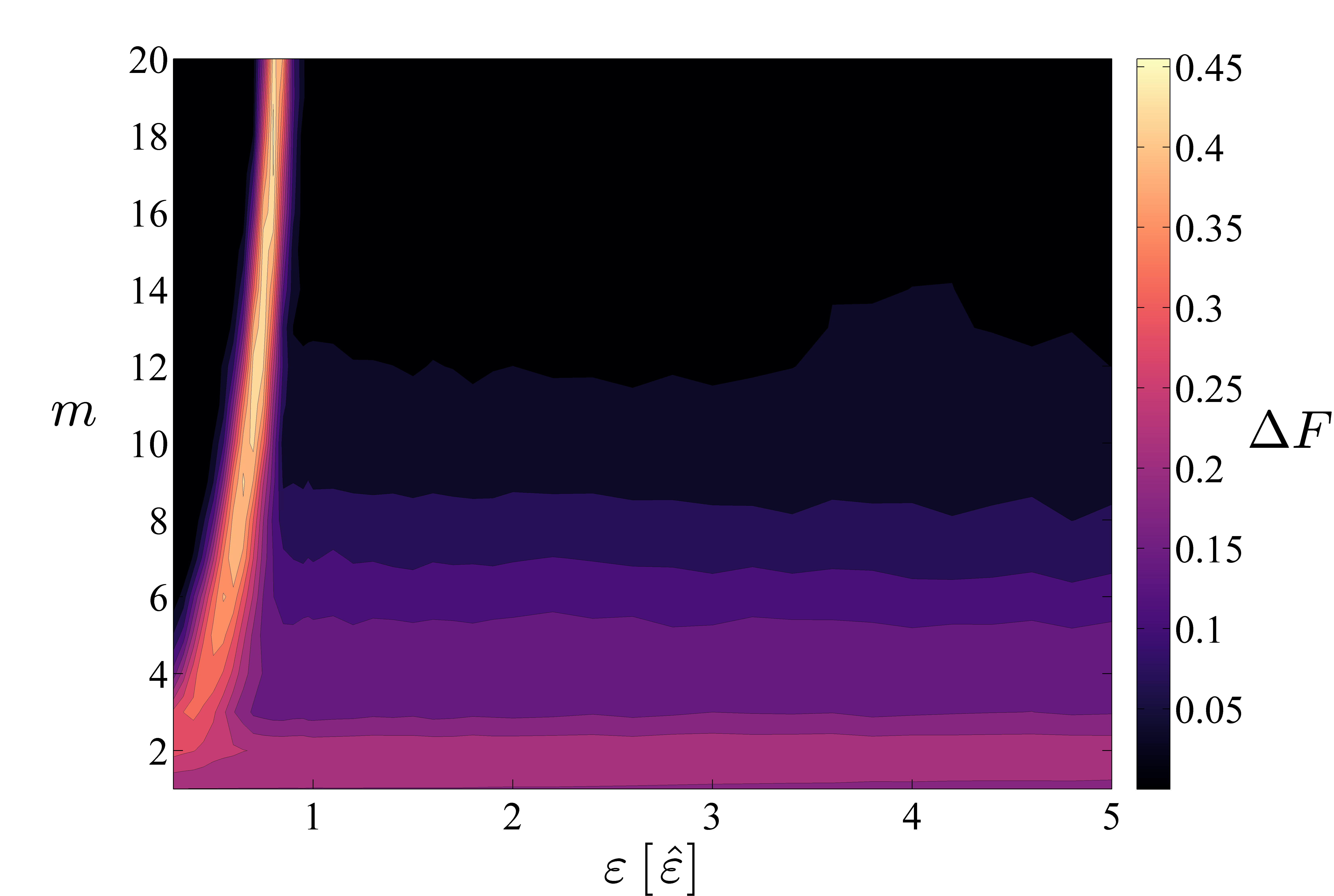}}
\end{center}
\end{minipage}
\caption{\label{fig:cs}Fidelity $F(\hat{\varrho}_\mathrm{CS},\hat{\varrho}(m,\varepsilon))$ depending on the number of measurement settings $m$ and the model parameter $\varepsilon$ (top) and corresponding standard deviation $\Delta F$ (bottom) obtained via bootstrapping. Since in compressed sensing we are more interested in the regime of few measurement settings and the fidelities do not change significantly for larger $m$, we restrict ourselves to the region with $m\leq 20$. The data are generated randomly from $\hat{\varrho}_\mathrm{CS}$ and the measurement settings per $m$ are drawn randomly as well. The fidelities are averaged over different data realisations and measurement settings. The highest fidelities are achieved for $\varepsilon\approx \hat \varepsilon$ with rapid decrease for $\varepsilon<\hat \varepsilon$ where the fraction of infeasible optimisations increases. Note that the higher the fidelity, the lower the standard deviation. }
\end{figure}

The results for varying $m$ and $\varepsilon$ in units of $\hat \varepsilon$ are shown in Fig.~\ref{fig:cs}. We compare the reconstructed states to $\hat{\varrho}_\mathrm{CS}$, which we used to generate the simulated data. We see that as $m$ increases, the fidelity converges to unity at $\varepsilon=\hat{\varepsilon}$ (where $\hat{\varrho}_\mathrm{CS}$ is defined). We are interested in how quickly our reconstructed state approaches the optimal $\hat{\varrho}_\mathrm{CS}$ with fewer measurement settings, particularly if $\varepsilon$ is misestimated. For instance, we see that we can obtain average fidelities of more than $0.8$ for only 6 measurement settings. Fig.~\ref{fig:cs} (top) again illustrates that $\varepsilon=\hat{\varepsilon}$ is the best choice as the fidelities around this region (and away from pathologically small numbers of measurement settings, $m>3$) are the highest. Moreover, we also see that with increasing $m$, the standard deviation $\Delta F$ of the fidelity becomes smaller for $\varepsilon\geq\hat \varepsilon$. For $\varepsilon<\hat \varepsilon$, infeasibilities of the optimisation Eq.~\eqref{eq:opt} that appear for certain choices of measurement settings lead to large standard deviations, which can be seen by the ridge in the area left of $\varepsilon=\hat \varepsilon$ in Fig.~\ref{fig:cs} (bottom). The ridge as well as the region of infeasibility gets close to $\varepsilon=\hat \varepsilon$ for large $m$, which is reasonable since more information (i.e. more constraints) puts greater restrictions on the optimisation problems. If fewer measurement settings are considered, as in the highly tomographically incomplete regime, overestimation of $\varepsilon$ is less detrimental and state estimates still perform well, i.e. the fidelity is relatively constant for $\hat \varepsilon \lesssim \varepsilon \lesssim 3 \, \hat \varepsilon$. However, as $m$ increases, the reconstruction becomes more strongly dependant on the choice of $\varepsilon$. Generally, we see that the higher the fidelity, the lower the standard deviation.

\subsection*{Discussion}\label{sec:results}

In this work, we have experimentally explored the compressed sensing paradigm for quantum state tomography as applied to the photonic setting. We have explicitly laid out a method for applying these techniques and reconstructed the state of a four-photon system with  tomographically complete data available, observing a high fidelity of the reconstructed state with respect to the target state.
The presence of noise in the data requires that one carefully chooses appropriate constraints on the optimisation. In current applications, these parameters are usually obtained in an ad hoc way. We have provided a prescription to establish the parameters in a more systematic way by modelling the noise and performing cross validation, which is a general method for model selection.
The quality of the data, being afflicted with noise predominantly attributed to finite counting statistics, allows us to model the noise via a multinomial distribution. This is a situation commonly expected for photonic experiments with postselected data. In fact, we observe a great agreement between estimating the model parameter from theoretical noise modelling and cross validation.

Having established the appropriate model, we have been able to perform state reconstruction with tomographically incomplete data, which rapidly converges to the highest fidelity estimate as the number of measurement settings increases. As a validity check, we have also run different estimators on the full data and obtained similar results, showing that our compressed sensing procedure yields reasonable estimates. As is predicted by 
the mathematical theory of compressed sensing, we have found that the number of measurement settings needed for a satisfactory estimate of the underlying state is much smaller than the number of measurements necessary for tomographic completeness. We 
have also carried out a comprehensive bootstrapping analysis to build confidence in the robustness of our method. In fact, we 
have observed that the uncertainty in the fidelity quickly decreases with increasing number of measurement settings. 

Furthermore, we have studied the robustness of our method with respect to improper model selection and the effects on the reconstruction. We have found that for several choices of models and different numbers of measurement settings, the performance of the reconstruction can vary dramatically. For small numbers of measurement settings, our method depends less strongly on the model. In contrast, for large numbers of measurement settings, it is imperative to determine the appropriate model for optimal performance. 

These results confirm that compressed sensing in conjunction with 
suitable model selection gives rise to reliable procedures for state reconstruction leading to effective tomography with tomographically incomplete data. These techniques can be applied to a wide range of experimental settings and provide a means to identify and verify appropriate models thereby paving the way for the future of practical quantum state tomography.
With this, we contribute to establishing compressed sensing as a practical tool for quantum state tomography in the low-information regime.


\section*{Acknowledgements}
We thank the DFG (EI 519/7-1, EI 519/9-1 within SPP 1798 CoSIP), the Templeton Foundation, the EU (AQuS, SIQS, RAQUEL), EU FP7 grant 600838 QWAD, U.S. Army Research Office (ARO) grant W911NF-14-1-0133, the BMBF (Q.com), the Freie Universi{\"a}t Berlin within the Excellence Initiative of the German Research Foundation, the South African Research Foundation, the Fritz Haber Institute of the Max Planck Society, and the German National Academic Foundation (Studienstiftung des Deutschen Volkes) for support.

\bibliographystyle{unsrt}

\appendix

\section{Fidelity estimation with error bound}\label{appx:FidelityEstimation}

In this section, we provide more detail to the fidelity estimation with an error bound without the need of resorting to quantum state tomography.
In the Pauli-operator basis
\begin{equation}
\{O_l : \  O_l \in \bigotimes_{j=1}^n \{\mathbbm{1}, \sigma_x, \sigma_y, \sigma_z\} \},
\end{equation}
 we can estimate from the measured probabilities $\hat{p}_{j,k} = y_{j,k}/N_j$ the expansion coefficients
\begin{equation}
	\xi^l_\varrho = \tr\left(\varrho\, O_l/\sqrt{d}\right)
\end{equation}
of the prepared state $\varrho$  by a linear transformation $\Omega$,
\begin{equation}
	\pmb{\xi}_\varrho = \Omega\ \hat{\pmb{p}}.
\end{equation}
For convenience, we denote by $\hat{\pmb{p}}$ 
the row-vectorisation of the matrix with entries $\hat{p}_{j,k}$.
The fidelity with respect to a pure target state $\varrho_\mathrm{T}$ can be written in terms of the expansion coeffcients as  
\begin{align}
	F^2(\varrho_{\text{T}},\varrho) &= \sum_l \xi^l_{ \varrho_{\text{T}}} \xi^l_\varrho  \nonumber \\ \label{eq:app:fidelity}
		&= \pmb{\xi}^T_{ \varrho_{\text{T}}} \Omega\ \hat{\pmb{p}}.
\end{align}
The frequency of the $d$ different outcomes for the $j$-th measurement setting is described by a multinomial distribution. 
The covariance matrix is given for each multinomial distribution by 
\begin{equation}
	\operatorname{Cov}(Y_{j,k}, Y_{j,l}) = N_j \left( p_{j,k} \,\delta_{i,j} - p_{j,k}\, p_{j,l}\right).
\end{equation}
Since different measurement settings correspond to mutually orthonormal operators, the frequencies of different measurement settings are uncorrelated, i.e.\ $\operatorname{Cov}(Y_{i,k}, Y_{j,l}) = 0$ for~$i\neq j$. 
Therefore the covariance matrix for the probabilities $\hat{\pmb{p}}$ can be estimated from the data as
\begin{equation}
	\operatorname{Cov}(\hat{p}_{j,k}, \hat{p}_{j,l}) = N_j^{-1} \left( \hat{p}_{j,k} \delta_{i,j} - \hat{p}_{j,k} \hat{p}_{j,l}\right).
\end{equation}
By means of linear error propagation, the variance of the fidelity is given by 
\begin{equation}
	\operatorname{Var}(F^2) = \pmb{\xi}^T_{ \varrho_{\text{T}}}\, \Omega \, \operatorname{Cov}(\hat{p},\hat{p})\, \Omega^T\, \pmb{\xi}_{ \varrho_{\text{T}}},
\end{equation}
which yields an estimate of the statistical error of the fidelity estimate from the data
\begin{equation}
	\Delta F^2(\varrho, \varrho_{\text{T}}) = \sqrt{\operatorname{Var}(F^2)}.
\end{equation}
In particular, in order to estimate the fidelity with respect to the GHZ state, only nine Pauli basis measurements contribute. This can be seen from the expansion of the GHZ density matrix in the Pauli-operator basis 
\begin{equation}\label{eq:app:expansion}
	\begin{split}
	\varrho_\text{GHZ} = \frac{1}{16} \biggl[\ \sum_{\sigma \in \{\mathbbm{1}, \sigma_x, \sigma_y, \sigma_z\}}\hspace{-.5cm} \sigma^{\otimes 4} &+ \sum_\text{Perm.} \mathbbm{1} \otimes \mathbbm{1} \otimes  \sigma_z \otimes \sigma_z \\ &+ \sum_\text{Perm.} \sigma_x \otimes \sigma_x \otimes \sigma_y \otimes \sigma_y 
	\biggr],
	\end{split}
\end{equation}
where the last two sums run over all six distinct orders of the four factors of the tensor product. 

To estimate the fidelity \eqref{eq:app:fidelity}, only the $16$ Pauli coefficients of the prepared state are required that correspond to the operators of the expansion~\eqref{eq:app:expansion}. From the measurement outcomes of the measurement setting $\sigma_z^{\otimes 4}$, all coefficients of operators containing only the identity $\mathbbm{1}$ and $\sigma_z$ can be estimated. Thus, only nine Pauli basis measurements are necessary to estimate the fidelity. 

Note that it is also possible to employ the measurement outcomes of all other measurement settings in the estimation of coefficients of terms that include the identity in Eq.~\eqref{eq:app:expansion}. In principle, it is thereby possible to further reduce the statistical error of the estimate of those coefficients. However, for the data set considered in this work, using more than nine measurement settings does not significantly alter the fidelity estimate.

\end{document}